# Optomechanical properties of GaAs/AlAs micropillar resonators operating in the 18 GHz range

F. R. Lamberti,[1,2,#] Q. Yao, [2,#] L. Lanco,[1,3] D. T. Nguyen,[2] M. Esmann,[1] A. Fainstein,[4] P. Sesin,[4] S. Anguiano,[4] V. Villafañe,[4] A. Bruchhausen,[4] P. Senellart,[1] I. Favero,[2] N. D. Lanzillotti-Kimura[1,*]

[1] Centre de Nanosciences et de Nanotechnologies, CNRS, Univ. Paris-Sud, Université Paris-Saclay, C2N – Marcoussis, 91460 Marcoussis, France

[2] Laboratoire Matériaux et Phénomènes Quantiques, CNRS UMR 7162, Université Paris Diderot, Sorbonne Paris Cité, 75013 Paris, France

[3] Département de Physique, Université Paris Diderot, Sorbonne Paris Cité, 75013 Paris, France

[4] Instituto Balseiro & Centro Atomico Bariloche, CNEA, 8400 Rio Negro, Argentina

\# Equally contributors
\* Corresponding author: daniel.kimura@c2n.upsaclay.fr

## Abstract

Recent experiments demonstrated that GaAs/AlAs based micropillar cavities are promising systems for quantum optomechanics, allowing the simultaneous three-dimensional confinement of near-infrared photons and acoustic phonons in the 18-100 GHz range. Here, we investigate through numerical simulations the optomechanical properties of this new platform. We evidence how the Poisson's ratio and semiconductor/vacuum boundary conditions lead to very distinct features in the mechanical and optical three dimensional confinement. We find a strong dependence of the mechanical quality factor and strain distribution on the micropillar radius, in great contrast to what is predicted and observed in the optical domain. The derived optomechanical coupling constants $g_0$ reach ultra-large values in the $10^6$ rad/s range.

The study of mechanical systems in their quantum ground state motivates the development of novel optomechanical resonators with frequencies higher than a few GHz.[1–4] In this particular frequency range, standard cryogenic techniques become sufficient to reach the quantum regime without relying on additional sideband optical cooling. Recently, GaAs/AlAs pillar microcavities have been presented as new optomechanical resonators performing in the unprecedented 18-100 GHz mechanical frequency range, showing highly promising features such as state-of-the-art quality factor-frequency products.[5] Well known for their optical properties, micropillar cavities confine light in the three directions of space. They are widely used in non-linear optics, taking advantage of the strong optical non-linearities in GaAlAs semiconductors,[6–8] in optical simulations based on quantum well cavity polaritons,[9–12] and in solid state quantum optics where single quantum dots constitute highly coherent artificial atoms.[13,14] This diversity of optical applications opens a wide range of possibilities in the field of optomechanics, such as the creation of nonclassical and entangled photonic and mechanical states,[1] and the development of hybrid quantum devices that interface usually incompatible degrees of freedom by means of phonons.[15,16] Their properties as optomechanical resonators thus need to be explored to determine the acoustic confinement mechanism, and the optimal optomechanical coupling conditions. Mechanical micropillars were previously studied both theoretically and experimentally.[5,17,18] However, the detailed description of the mechanical eigenmodes, the corresponding extremely high optomechanical coupling factors, and the optomechanical interactions with quantum emitters were not addressed before. In this Letter, we report such a comprehensive theoretical study. In particular, our work reveals that while the mechanism governing the acoustic confinement along the pillar axis is very similar to the optical confinement, the Poisson's ratio (i.e. the signed ratio of transverse elastic strain to axial elastic strain) and a different set of boundary conditions lead to very distinct acoustic/optical confinement in the transverse direction. The optomechanical coupling factors reach values of approximately $10^6$ rad/s and the mechanical Q-factors exceed $10^3$, revealing the potential of micropillars as optomechanical resonators. In addition, we report on the conditions to optimize the coupling of quantum emitters to this optomechanical platform.

An efficient optomechanical transducer requires a good overlap between a confined optical and a confined acoustic mode.[19–21] In a semiconductor pillar microcavity, confinement of optical fields is achieved by enclosing a $m\lambda/2$ thick cylindrical layer in between two highly reflective distributed Bragg reflectors (DBRs), following the concept of a Fabry-Perot resonator.[22,23] Here, $\lambda$ is the optical wavelength in the spacer material and $m$ an integer number. DBRs are obtained from stacking $\lambda/4$ layers of materials with contrasting indices of refraction. In the lateral direction the contrast in the index of refraction between the pillar and the surrounding medium induces a confinement of the electromagnetic field, as for an optical fiber. This confinement, in addition to the vertical confinement provided by the DBRs, allows reaching small optical mode volumes and high optical quality factors, important for cavity quantum electrodynamics and optomechanical applications.

Following the same concept, an acoustic semiconductor pillar microcavity (see the schematics in Fig. 1a) can be formed by enclosing a $n\lambda/2$ thick cylindrical layer in between two highly reflective acoustic DBRs, where $\lambda$ is the phonon wavelength in the spacer material and $n$ an integer number.[5,17] In this case, DBRs are obtained by stacking layers of materials with contrasting acoustic impedances, determined by the elastic properties of the materials.[19,24] For the particular material choice of GaAs/AlAs the optical and acoustic impedance contrasts are almost equal.[19,25] Hence, the micropillar is an optical resonator, confining photons in the near infrared range (~ 325 THz) and, simultaneously a mechanical resonator, confining acoustic phonons of around 18 GHz, both photons and phonons having the same wavelength $\lambda$ ~ 250 nm in GaAs.[5] There are, however, two important differences between the optical and mechanical behaviors of the system. Firstly, phonons are subject to total reflection due to the incapability of the semiconductor/vacuum interfaces to transmit mechanical vibrations and secondly, different mechanical directions of vibration are coupled through the Poisson's ratio. As discussed hereafter, both effects have an important impact on the quality factor, mechanical field distribution and vacuum optomechanical coupling factors.

To evidence these optomechanical properties, we simulate the behavior of a 3 μm diameter micropillar formed by 2 DBRs enclosing a $\lambda/2$ GaAs spacer. Each DBR is formed by 10 periods of GaAs/AlAs ($\lambda/4$, $\lambda/4$). The structure is surrounded by vacuum, placed on top of a GaAs substrate, unless indicated otherwise. In Fig. 1a we show the schematics of such GaAs/AlAs pillar microcavity. We performed the calculation of the electromagnetic and mechanical eigenmodes using a commercial finite elements method software (COMSOL). The numerical model uses a 2 dimensional axisymmetric geometry. GaAs and AlAs are considered here as isotropic materials, both for optical and mechanical properties. No optical or mechanical absorption processes are considered. For the mechanical calculations a free boundary condition simulates the sample-vacuum interfaces. Perfectly matched layers (PMLs) were used to simulate the infinite GaAs substrate. For the optical simulations, we evaluate the indices of refraction of GaAs and AlAs by using the Afromowitz method.[26] PMLs were also implemented to simulate the infinite vacuum environment surrounding the structure.

In one-dimensional systems, the optical wave equation can be perfectly mapped into the elastic wave equation, where the indices of refraction play the role of the acoustic impedances. As a consequence, in a one-dimensional GaAs/AlAs planar cavity for which optical and acoustic impedance contrasts are almost the same, the displacement pattern of the mechanical confined mode perfectly overlaps the electric field spatial profile of the optical confined mode.[19,25,27,28] The electric field distribution inside a 3D micropillar structure is presented in Fig. 1b, corresponding to the confined optical mode with $\lambda = 921$ nm. The confinement of the electric field and thus its modulation in the vertical direction replicates the behavior observed in a planar structure. In a cavity with a $\lambda/2$ GaAs spacer the modulus of the electric field $|\vec{E}|^2$ presents a minimum at its center along the vertical direction. In the radial direction a smooth Bessel function distribution of the intensity reflects the axial symmetry of the system. In Fig. 1c we show the acoustic displacement $|\overrightarrow{u(r,z)}|^2$ distribution corresponding to a confined phononic mode around 18.2 GHz, and Fig. 1d presents the corresponding volumetric strain distribution $\left|\frac{\Delta V}{V}\right|$. We observe that for both Fig. 1c and Fig. 1d, the mechanical quantities plotted present the same Bessel-like radial envelope, similar to the one shown in Fig. 1b for the electric field

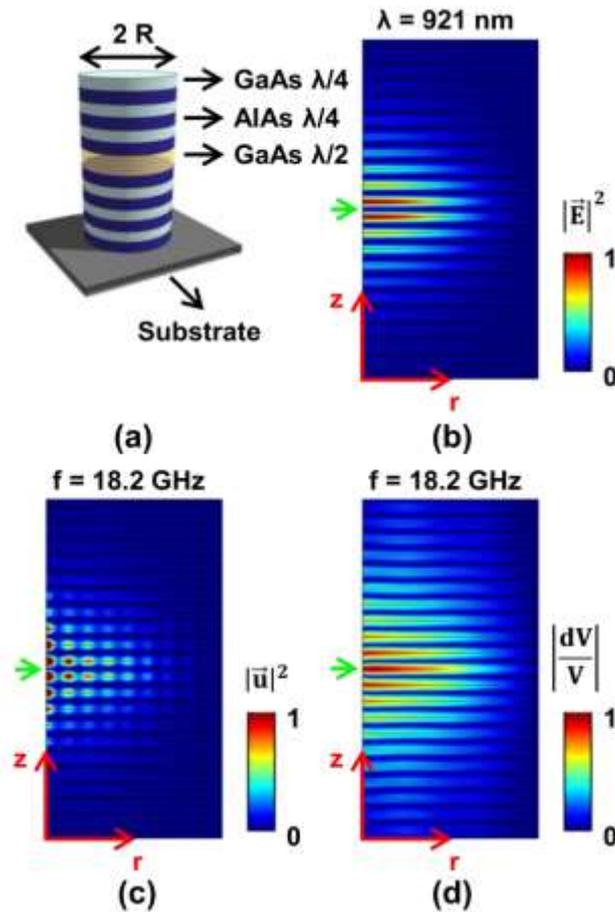

Figure 1: Optomechanical pillar microcavity. (a) Schematics of the micropillar structure, formed by two distributed Bragg reflectors (DBR) embedding a spacer. The number of periods in the shown DBR is arbitrary. (b) Modulus square of the electric field distribution in a 10 periods/DBR resonator with a $\lambda/2$ cavity, for a mode of 921 nm of optical wavelength. (c) Modulus square of the total mechanical displacement distribution corresponding to the confined acoustic mode at a resonance frequency of 18.2 GHz. Note the vertical modulation matching the one shown in (b), in addition to the radial modulation due to the free surface boundary condition. (d) Modulus of the volumetric strain. For these numerical simulations, the micropillar is considered without any GaAs substrate.

In contrast with what is observed for the electric field, the mechanical displacement profile $|\vec{u}(r,z)|^2$ presents a strong radial modulation, characteristic of the 3D micropillars. As we show below the presence of the resulting nodes can be explained as a consequence of the boundary conditions and the coupling between different mechanical degrees of freedom induced by the Poisson's ratio. Moreover, as will also be shown, there is a clear difference between the frequency dependence of acoustic and optical modes as a function of the radial size of the micropillar. Indeed, on the optical side, a reduction of the micropillar radius induces a smooth increment in the frequency of the confined optical mode. In Fig. 2a we show the calculated frequency of the optical eigenmode as a function of the radius. A monotonic dependence is observed, which asymptotically approaches the eigenfrequency of the mode of the planar cavity. For comparison, in Fig 2b, the frequency of the mechanical eigenmode is presented as a function of the micropillar radius. Several branches are apparent, evidencing a series of anticrossings, in strong contrast with the case of light confinement. In red, a guide to the eye is included indicating the frequency dependence of the center of the branches, showing a trend similar to the one observed in the optical case.

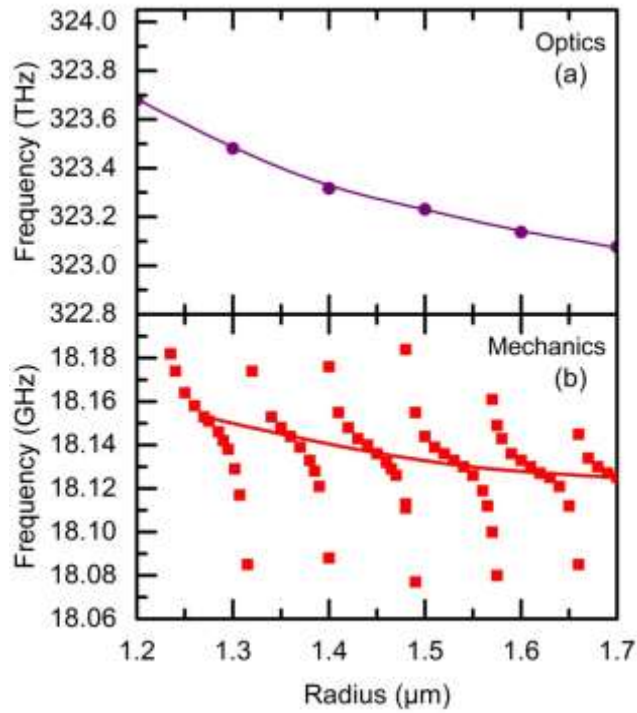

Figure 2: Dependence of the fundamental confined optical and acoustic mode frequencies on the radius of the micropillar. (a) Optical dependence showing a monotonic increase of the frequency as the radius is reduced. (b) Multiple anticrossings in the mechanical dependence originating from the coupling of vertical and radial modes due to the Poisson's ratio. The red line is a guide to the eye to show the dependence of the frequency with the radius of the micropillars, at the center of the branches.

To understand the origin of these effects we consider a simplified system, a uniform solid GaAs cylinder. In this case, the vertical acoustic confinement is given not by a DBR butic by a free GaAs/vacuum interface. The tensorial stress components (in cylindrical coordinates) $\sigma_{ij}$ must then satisfy the following simultaneous boundary conditions

$$\sigma_{zz} = \sigma_{z\varphi} = \sigma_{zr} = 0 \quad \text{at } z=0 \text{ and } z=2H$$
$$\sigma_{rr} = \sigma_{r\varphi} = \sigma_{rz} = 0 \quad \text{at the cylindrical surface } (r=R).$$

where 2H is the height of the cylinder and R the micropillar radius. The Poisson's ratio $\nu$ of a material establishes a direct relationship between transverse and axial strains.

For the particular case $\nu = 0$, the displacements ($u$) that comply with this set of boundary conditions can be expressed as

$$u_{z,k} \propto \cos(kz)$$
$$u_{r,q} \propto qJ_1(qr)$$

Here, $J_1$ is the Bessel function of the first kind of order 1, k is a wavenumber defined along the vertical direction, q an in-plane wavenumber and z and r the vertical and radial coordinates. These two solutions are also those of an infinite planar structure (cylinder of infinite radius), and of an infinitely long cylinder, respectively. In the case of an infinitely long cylinder, the displacement pattern presents a radial modulation due to the free surface boundary conditions at r=R. The number of nodes in the radial direction is set by the value of the wavenumber q in $J_1$. In the case of an infinite planar structure, a vertical modulation is also present due to the free surface boundary condition, but in the vertical direction (for z = 0 and z = 2H). The number of nodes is in this case determined by the value of the wavenumber k.

Fig. 3 presents the mechanical eigenfrequencies in a GaAs cylinder for very small radii and for two cases: $\nu = 0$ and $\nu = 0.31$ (the real value for GaAs). The displacement profiles of the modes are shown as intensity maps in the insets. In the case $\nu = 0$, we recover two uncoupled modes: a vertical mode with a radial displacement $u_r = 0$ (blue curve) and a radial mode with a vertical displacement $u_z = 0$ (green curve). The frequency of the vertical mode is independent of the radius of the micropillar, and its displacement profile (intensity map A) is constant along the radial direction. The radial mode presents an increasing frequency when reducing the radius. Its displacement profile is constant along the vertical direction (intensity map B).

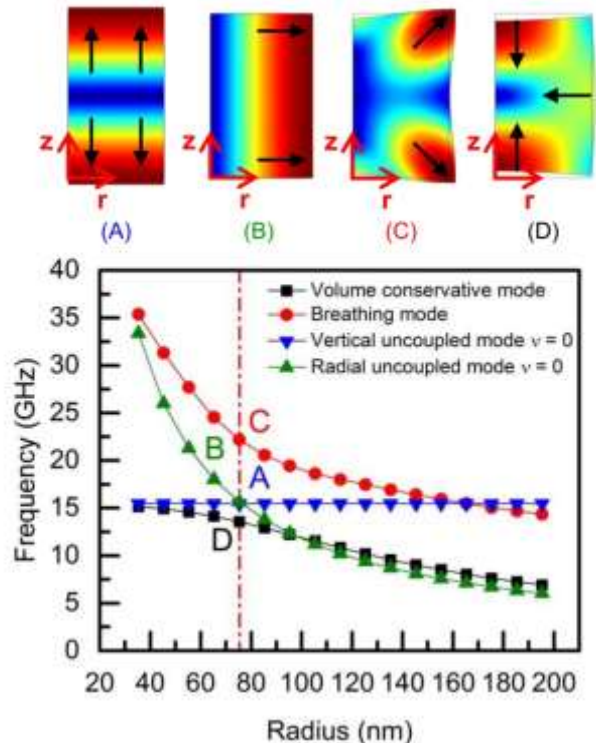

Figure 3: Dependence of the mechanical modes frequencies on the radius of a uniform GaAs cylinder. Curves A and B correspond to the vertical and radial modes, respectively, when $\nu = 0$. Curves C and D correspond to the coupled hybrid vertical-radial modes ($\nu \neq 0$). Case C (D) is a breathing (volume conservative) mode. The displacement profiles are plotted for a structure of 75 nm of radius.

When considering a Poisson's ratio different from zero (ν = 0.31), a mode mixing takes place, and two new modes appear. Each one of these two modes presents simultaneously vertical and radial displacement components. The red curve corresponds to a breathing mode (intensity map C), where a vertical expansion is in phase with a radial expansion. The black curve is associated to a volume-conservative mode (D). An anticrossing between the red and black curves around 75 nm of radius can be noted. Far from the anticrossing the mixed modes tend to mimic the behavior of the vertical or radial modes.

For a given pillar mechanical eigenmode of frequency $\omega_m$, a Poisson's ratio ν ≠ 0 implies that a vertical strain coexists with a nonzero radial strain. These components are respectively subject to the vertical and radial boundary conditions. In the case of a micropillar, a mode has to fulfill two simultaneous resonance conditions: the vertical confinement, in this case determined mainly by the DBRs and the spacer thickness, and the radial confinement determined by a zero stress condition (normal to the surface) at the vacuum/sample interface at r = R. Fulfilling both conditions results in a coupling between vertical and radial strains described by the Poisson's ratio. This leads to simultaneous generation of nodes in the displacement pattern in both the vertical and radial direction (Fig. 1c). Figure 4a presents a detail of the dependence of the acoustic mode frequency with the micropillar radius around 1.53 um. We point out that the eigenfrequencies do not vary with the number of pairs in the DBRs. The fraction of mechanical energy α stored in the radial direction can be determined as:

$$\alpha = \frac{\int_{pillar} \frac{1}{2} \times \omega^2 \times |u_r|^2 \times \rho \times dV}{\int_{pillar} \frac{1}{2} \times \omega^2 \times |u|^2 \times \rho \times dV}$$

where ω is the eigenfrequency of the mode, ρ the density of the materials, $|u_r|$ the modulus of the displacement radial component, and $|u|$ the modulus of the total displacement. This quantity is evaluated in Fig. 4b for two structures with 10 and 20 AlAs/GaAs pairs in each DBR. For radii where the eigenfrequencies remain almost constant the energy is mainly stored with a vertical displacement profile, whereas in the edges of the branch plotted in Figure 4a the modes present displacements with a mixture of vertical and radial components.

The mechanical Q-factor quantifies how well a resonator is mechanically isolated from its environment, in particular from the substrate. The imaginary part of the eigenfrequency gives access to the mechanical quality factor of the micropillar Q = Re($\omega_m$)/2Im($\omega_m$). The dependence of the Q-factors on the radius is shown in Fig. 4c-d for the considered branch, and for two structures with 10 and 20 pairs on each DBR, respectively. In Fig. 4c we observe a decrease of the Q-factor when approaching the vertical-like modes, while in Fig. 4d we observe an increase of the Q-factor when approaching this region.

This behavior can be explained by considering independently the coupling to the substrate of the radial and vertical components. On one hand, the coupling of a vertical mode to the substrate –and thus the Q-factor- strongly depends on the reflectivity of the bottom DBR. On the other hand, the losses of the radial modes we can infer less sensitive to the reflectivity of the DBR. In the case of a DBR formed by 10 periods (Fig. 4c), the Q-factor associated to the radial mode is higher than the one associated to the vertical mode. As such, a vertical-like mode at the center of the branch presents a lower Q-factor than a mixed radial-vertical mode. Conversely, in a cavity with 20 period DBRs, the Q-factor of the vertical mode is higher than the one of the radial mode. As such, the Q-factor in the central part of the branch shown in Fig. 4d presents a maximum, where the mode remains mainly vertical. By keeping the same number of periods in the DBRs, a strong dependence of the Q-factor on the radius is observed, which can give rise to changes by a factor of 4 within a 100 nm radius range.

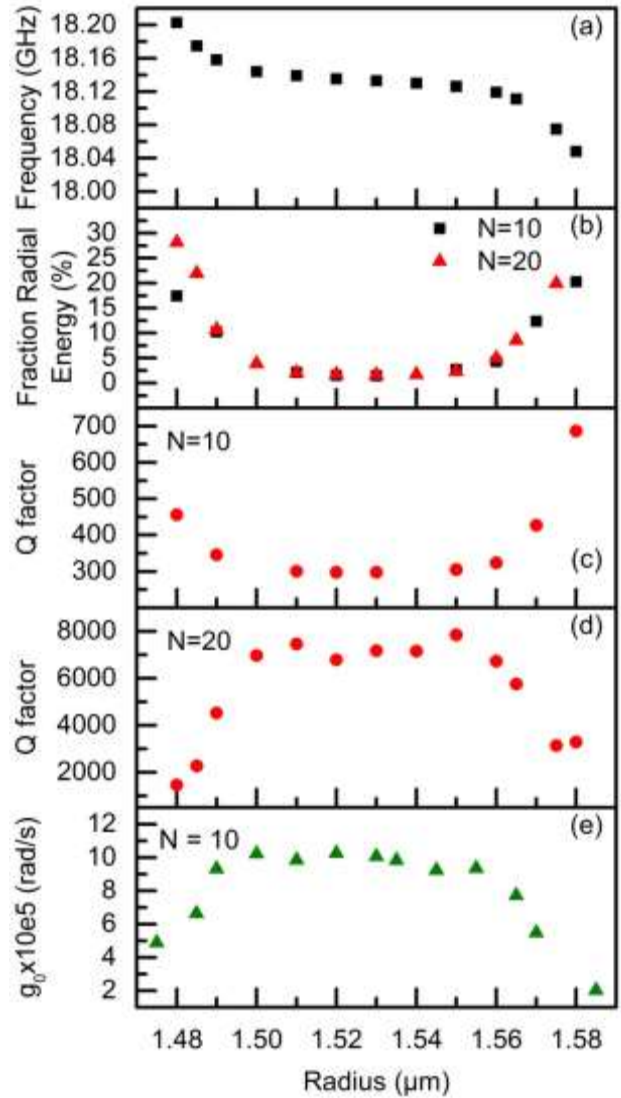

Figure 4: (a) Dependence of the fundamental mechanical frequency of a micropillar with the radius. (b) Fraction of radial energy in the mechanical eigenmode as a function of the radius for 10 and 20 pairs (N) of GaAs/AlAs in the DBRs. (c) and (d) plot the Q-factor as a function of the radius for N = 10 and N = 20. For the 20 periods case a maximum Q-factor of ~8000 is reached. (e) Optomechanical coupling $g_0$ as function of the radius, reaching a maximum of about $10^6$ rad/s. Notice the strong dependence of these magnitudes near the extremes of the plotted branch.

In order to determine the potential of these structures as optomechanical resonators, we calculated the vacuum optomechanical coupling factor $g_0$ by considering two parts: the geometrical coupling factor $g_0^{geom}$ and the photoelastic coupling factor $g_0^{photo}$.[19,29,30] The first term describes coupling due to changes of the microcavity shape and dimensions, whereas the second corresponds to changes of the material indices of refraction produced by mechanical strain. We let AlAs parts aside and consider only the photoelastic effect in GaAs, which is estimated to be the dominant contribution for energies near the electronic bandgap of GaAs. The only components of the photoelastic tensor affecting the GaAs dielectric tensor in our model are $p_{11}$ and $p_{12}$. The value of these components at an optical wavelength of 920 nm are $p_{11} = 0.276$ and $p_{12} = 0.305$.[19,31] In Fig. 4e we report the calculated $g_0$ for the case of a cavity formed by a 10 periods DBR on each side. The change in the acoustic field distribution inside the micropillar as a function of radius modifies the overlap with the optical field distribution. This also leads to a strong dependence of the optomechanical coupling on the radius, as seen in Fig. 4e. We observe that in the flat regions of the branch shown in Fig. 4a, where the displacement is mainly in the vertical direction, the $g_0$ reaches values of the order of $10^6$ rad/s, comparable with state-of-the-art nano-optomechanical devices.[30,32] In spite of the reported strong radial modulation of displacement, the electric field distribution of a confined mode presents a very good spatial overlap with the displacement field, resulting in large $g_0$ values. Note, however, that $g_0$ drops by one order of magnitude close to the edges of the branches.

The $g_0$ results from the modification of the indices of refraction and the geometrical deformation of the interfaces of the micropillar structure, i.e. on the overall overlap between the confined optical and mechanical mode. The detailed description of the acoustic eigenmodes is also particularly relevant when a local coupling should be considered. For instance, a single quantum emitter *locally* interacts with the optical and strain fields. As shown in Fig. 1d, in a $\lambda/2$ cavity there is a maximum of the mechanical strain at the center of the spacer. At this point, there is a node of the optical field. Current nanofabrication techniques allow to precisely position quantum emitters in a micropillar. By vertically positioning a quantum emitter away from the center of the spacer it is thus possible to reach a situation where it probes both the confined optical and mechanical fields. This would enable the study of tripartite systems where the interactions between confined ultra-high frequency acoustic phonons, optical photons and a 2-level system can be deterministically engineered.

In conclusion, we have investigated the main characteristics of GaAs/AlAs micropillar optomechanical resonators, unveiling a feature-rich mechanical response of the structures. These resonators work at unprecedented high mechanical frequencies (around 18 GHz), with high quality factors (greater than $10^3$ for the considered DBR configurations), and in addition can provide very high optomechanical coupling factors (of the order of $10^6$ rad/s). Due to the coupling of the radial and vertical mechanical degrees of freedom in micropillars the spatial strain profiles of the confined modes strongly depend on the micropillar radius. This dependence entails a strong variation of the optomechanical coupling with the radius of the micropillar and therefore has to be taken into account in the design of micropillar-based optomechanical resonators. Finally, a full understanding of the tridimensional confinement of acoustic phonons in micropillars enables a completely novel feature to the quantum information toolbox: the possibility of actually engineering the phononic landscape of nanostructures and controlling the acoustic-phonon interactions with other solid state excitations.

**Acknowledgments.**

Work partially supported by a public grant overseen by the French National Research Agency (ANR) as part of the "Investissements d'Avenir" program (LabexNanoSaclay, reference: ANR-10-LABX- 0035), the ERC Starting Grants No. 277885 QD-CQED, No. 715939 NanoPhennec and No. 306664 Ganoms, the French Agence Nationale pour la Recherche (grant ANR QDOM), the French RENATECH network, the ANPCyT Grants PICT 2012-1661 and 2013-2047, and the international franco-argentinean laboratory LIFAN (CNRS-CONICET). N.D.L.K. was partially supported by the FP7 Marie Curie Fellowship OMSiQuD.